\documentstyle[prd,aps,psfig]{revtex}
\draft 
\begin{document}

\twocolumn[\hsize\textwidth\columnwidth\hsize\csname
@twocolumnfalse\endcsname

\def\boxit#1{\vbox{\noindent\hrule\hbox{\vrule\hskip 3pt
\vbox{\vskip 3pt
\hbox{#1}\vskip 3pt}\hskip 3pt\vrule}\hrule}}

\title{Bounds on neutron-star moments of inertia and the evidence for
general relativistic frame dragging}

\author{Vassiliki Kalogera and Dimitrios Psaltis}

\address{Harvard-Smithsonian Center for Astrophysics, 60 Garden St.,
Cambridge, MA 02138}

\date{Received March 26, 1999; published}
\maketitle

 \begin{abstract}
 Recent X-ray variability observations of accreting neutron stars may
provide the first evidence for frame dragging effects around spinning
relativistic objects. Motivated by this possibility and its implications
for neutron-star structural properties, we calculate new optimal bounds on
the masses, radii, and moments of inertia of slowly rotating neutron stars
that show kilohertz quasi-periodic oscillations (QPOs). These bounds are
derived under minimal assumptions about the properties of matter at high
densities and therefore are largely independent of the unknown equation of
state. We further derive a semi-analytical upper bound on the neutron-star
moment of inertia without making any assumptions about the equation of
state of matter at any density. We use this upper bound to show that the
maximum possible nodal precession frequency of an inclined circular orbit
around a slowly spinning neutron star is $\nu_{\rm NP}\simeq 45.2(\nu_{\rm
s}/300$~Hz)~Hz, where $\nu_{\rm s}$ is the spin frequency of the neutron
star. We conclude that the nodal-precession interpretation of
low-frequency QPOs in accreting neutron-stars is inconsistent with their
inferred spin frequencies or the identification of the highest-frequency
QPO with that of a circular Keplerian orbit in the accretion disk.
 \end{abstract}


\vspace*{0.5cm}]

\section{INTRODUCTION}

Astrophysical observations of general relativistic phenomena are rare and
often confined to effects observable within our solar system~\cite{GRT}.
Among galactic sources, relativistic compact objects, such as neutron
stars and black holes, are the prime candidates for the detection of such
phenomena. In these sources, however, the effects of general relativity
are often convolved in complex ways with other astrophysical phenomena
making their identification often impossible. The observation, via precise
pulsar timing, of orbital shrinkage in double neutron-star systems
represents the most successful to date identification of pure general
relativistic effects in galactic compact objects~\cite{TW82}.

The recent discovery with the {\em Rossi X-ray Timing Explorer\/} of
quasi-periodic oscillations with variable kilohertz frequencies (hereafter
kHz QPOs) from many accreting neutron stars has provided us with a new
clock that measures accurately their variability~\cite{vdk98}. The
centroid frequencies of the highest-frequency kHz QPOs in each source have
been identified with Keplerian frequencies of stable orbits in the
accretion disks very close to the stars~\cite{vdk98,Setal96,MLP98}. These
QPOs can therefore be used as probes of the physical conditions in regions
where general relativistic effects in the strong-field regime are
non-negligible~\cite{MLP98}. This identification has lead to new upper
bounds on the masses (typically $\lesssim 2.2~M_\odot$) and radii
(typically $\lesssim 15$~km) of the accreting neutron
stars~\cite{MLP98,NSlim}. In one specific X-ray source (4U~1820$-$30), the
association of a kHz QPO with the Keplerian orbital frequency {\em at\/}
the radius of the innermost stable circular orbit around the neutron star
may provide us with the first evidence of this prediction of general
relativity as well as with the identification of the first relatively
massive ($\simeq 2.2~M_\odot$) neutron star~\cite{Zetal98}.

In many accreting neutron stars, a third low-frequency ($\simeq 10-70$~Hz)
QPO, the so-called horizontal-branch oscillation (HBO), is often observed
simultaneously with the kHz QPOs~\cite{vdK89,vdk98}. Soon after its
discovery the origin of this QPO was linked to the interaction between the
accretion disk and the magnetosphere of the neutron star~\cite{BFMA}.
Recently, however, this interpretation has been challenged~\cite{SV98} by
the understanding that the accretion disks appear to penetrate closer to
the stellar surface than what is required by the
model~\cite{vdk98,MLP98,Petal98}. As an alternative model, Stella \&
Vietri~\cite{SV98} suggested that the frequency of the HBO is the general
relativistic Lense-Thirring precession frequency, caused by frame
dragging, of an inclined circular orbit that has a Keplerian orbital
frequency equal to the frequency of the highest-frequency kHz QPO. The
physical mechanism responsible for producing brightness oscillations at
the precession frequency of a particular orbit in the accretion disk is
still a matter of active research~\cite{Mech}. However, when the effects
on the precession frequency of the quadrupole moment of the stellar
gravitation field can be neglected, the theoretically predicted
correlation between the precession and orbital frequencies is consistent
with observations~\cite{SV98,Petal98} (see Ref.~\cite{Petal98} for a
discussion of the discrepancy between the predicted and observed trend at
high frequencies). The normalization of the observed correlation, though,
is relatively high. In other words, the magnitude of the observed HBO
frequencies can be accounted by the model, {\em only if} the moments of
inertia of the neutron stars are $\simeq 4-5$ times larger than predicted
by any realistic equation of state~\cite{SV98,Petal98} (EOS). Therefore,
whether frame-dragging effects predicted by general relativity have
actually been observed in accreting neutron-stars depends crucially on our
knowledge of the moments of inertia of neutron stars.

Theoretical calculations of neutron-star properties such as their masses,
radii, and moments of inertia require the knowledge of the equation of
state of neutron-star matter up to densities $\gg 10^{15}$~g\,cm$^{-3}$.
However, the equation of state at densities higher than the nuclear
saturation density ($\simeq 2.7\times 10^{14}$~g\,cm$^{-3}$) is still
largely unknown and the subject of intense theoretical and experimental
research~\cite{G96}. Therefore accurate predictions for neutron-star
structural properties are hampered by the current uncertainty in the
equation of state of high-density neutron-star matter~\cite{CST94}.

In principle, the equation of state of matter at high densities could be
constrained by astrophysical measurements of neutron-star masses and
radii, although such measurements are relatively rare. The masses of the
compact objects in double neutron-star systems measured via precise pulsar
timing of the evolution of their binary orbits~\cite{TW82} were found to
cluster around $\simeq 1.35~M_\odot$~\cite{TC99}. Mass estimates of
neutron stars in other binary systems depend strongly on the unknown
inclination of the binary orbit and provide little additional information
on neutron-star properties~\cite{TC99}. Astrophysical measurements of the
radii of neutron stars based on the emitting area of their thermal
emission during thermonuclear bursts~\cite{LPT96}, during the quiescence
phase in transient systems~\cite{Retal99}, or during the cooling phase of
isolated neutron stars~\cite{PZ97} are also difficult because of the
systematic uncertainties in the predicted model spectra. Based on these
measurements so far no significant constraints were imposed on any of the
current equations of state.

Bypassing the uncertainties of the equation of state at high densities,
optimal bounds on the masses, radii, and moments of inertia of neutron
stars have been derived under minimal assumptions about the validity of
general relativity and the microscopic stability of neutron-star
matter~\cite{SH77}. Stricter bounds can be obtained if the so-called
causality limit is imposed, i.e., by requiring that the speed of sound is
less than the speed of light everywhere in the neutron star~\cite{RR74},
although the validity of this requirement has been questioned. Such limits
on the macroscopic properties of the neutron stars are largely independent
of their unknown equation of state.

In this paper, we obtain optimal bounds on the moments of inertia of the
neutron stars that show kHz QPOs under minimal or no assumptions for their
structure and equation of state, taking into account the bounds on their
masses and radii imposed by the observations of these QPOs. Our aim is to
address the possibility that low-frequency QPOs in neutron star sources
are related to general relativistic frame dragging which leads to
Lense-Thirring precession of inclined orbits.

In \S2, we discuss our assumptions and method of solution of the relevant
equations. In \S3, we present numerical and semi-analytical bounds on
neutron-star masses, radii, and moments of inertia and derive an
EOS-independent maximum limit on the (Lense-Thirring) nodal precession
frequency. In \S4, we discuss the implications of our results for the
Lense-Thirring interpretation of the observed HBOs in accreting neutron
stars.

\section{ASSUMPTIONS AND METHOD OF SOLUTION}

In the absence of detailed knowledge of the equation of state for matter
at very high densities ($\gg 10^{14}$~g~cm$^3$), we shall follow the
procedure outlined by Sabbadini \& Hartle~\cite{SH77} to obtain bounds on
the neutron-star masses, radii, and moments of inertia under a {\em set of
minimal assumptions} for the equation of state of neutron-star matter
above some fiducial energy density $\rho_0$. These are: (a)~the matter is
cold, (b)~the pressure in a given fluid element is determined uniquely by
its energy density, (c)~the energy density and pressure are everywhere
positive, and (d)~$dP/d\rho \ge 0$ (microscopic stability). In deriving
these new bounds we make use of the recent observations of kHz QPOs in
LMXBs to constrain the macroscopic properties of neutron stars in these
systems, i.e., their masses and radii~\cite{MLP98}.

The inferred spin frequencies of the neutron stars in the systems that
show kHz QPOs and HBO are $\simeq 250-350$~Hz~\cite{vdk98,Setal96,MLP98},
which are significantly smaller than the $\gtrsim 1500$~Hz breakup
frequency of a typical neutron star~\cite{CST94}. We, therefore, assume
that the neutron stars in all these systems are slowly rotating.
Hereafter, we also set $c=G=1$, where $c$ is the speed of light and $G$ is
the gravitational constant.

The frequencies of the kHz QPOs observed in the X-ray brightness of
accreting neutron stars and their dependence on mass-accretion rate have
lead to the identification of the highest-frequency QPOs with Keplerian
frequencies of stable circular orbits in the accretion
disks~\cite{vdk98,Setal96,MLP98}. In this interpretation, the observation
of kHz QPOs from an accreting neutron star imposes two constraints on its
mass and radius~\cite{MLP98}: (i) The radius of the orbit responsible for
the highest-frequency kHz QPO must be larger than the radius of the
innermost stable circular orbit around the neutron star. This leads to an
upper bound on the neutron-star mass:
 \begin{equation}
 M_{\rm NS}\le (\sqrt{864}\pi\nu_{\rm max})^{-1}{\rm ,}
 \label{Mmax}
 \end{equation}
 where $\nu_{\rm max}$ is the maximum observed frequency of the
highest-frequency QPO. (ii) The radius of the orbit responsible for the
highest-frequency kHz QPO must be larger than the radius of the
neutron-star itself. This leads to a mass-dependent upper limit on the
neutron star radius:
 \begin{equation}
 2 M_{\rm NS} \le R_{\rm NS} \le \left(
    \frac{M_{\rm NS}}{4\pi^2\nu_{\rm max}^2}
    \right)^{1/3}\;,
    \label{Rlim}
 \end{equation}
 where the lower limit on the neutron star is simply the requirement that
the central compact object is not a black hole. In writing
Eqs.~(\ref{Mmax}) and (\ref{Rlim}) we have neglected the fact that the
neutron star is slowly spinning; for the inferred spin frequencies of
neutron stars in these systems, the correction is only $\lesssim 10$\%
towards increasing these upper bounds.

In calculating the structure of a neutron-star with mass and radius
consistent with the above bounds, we divide it into two regions, given a
value of the fiducial density $\rho_0$, above which we do not trust the
equation of state: the core, with mass $M_{\rm c}$ and radius $R_{\rm c}$
in which $\rho\ge\rho_{\rm 0}$, and the envelope exterior to the core. It
has been shown~\cite{SH77} that the combined stellar configuration
satisfies the set of minimal assumptions for the equation of state for
neutron-star matter {\em if and only if\/}
 \begin{equation}
 M_{\rm c}\le \frac{2}{9}R_{\rm c}
  \left[1-6\pi R_{\rm c}^2 P_{\rm 0} +
     \left(1+6\pi R_{\rm c}^2 P_{\rm 0} 
   \right)^{1/2}\right]
 \label{Mcmax}
 \end{equation}
 and
 \begin{equation}
 M_{\rm c}\ge \frac{4\pi}{3} R_{\rm c}^3 
   \rho_{\rm 0}\;,
 \label{Mcmin}
 \end{equation}
 where $P_{\rm 0}$ is the pressure that corresponds to the energy density
$\rho_0$ at the edge of the core. 

Among all core configurations that satisfy the core mass
limits~(\ref{Mcmax}) and (\ref{Mcmin}), there is one that maximizes the
stellar moment of inertia~\cite{SH77}. The maximizing core configuration
is that of constant density $\rho_{\rm c} \geq \rho_{\rm 0}$. For a given
equation of state and fiducial energy density $\rho_{\rm 0}$, and for a
given neutron-star mass and radius we calculate the structure of the
neutron star for the core configurations that maximize its moment of
inertia. We then scan the ranges of neutron-star masses and radii allowed
by the kHz QPOs and obtain the global optimal bounds on the neutron-star
moment of inertia as a function of $\rho_{\rm 0}$.

To first order in the stellar angular velocity $\Omega$, the metric in and
around a slowly rotating star is~\cite{H67}
 \begin{eqnarray}
 ds^2&=&-e^{\nu(r)}dt^2
 +\frac{1}{1-2m(r)/r}dr^2
 +r^2(d\theta^2+\sin^2\theta d\phi^2)\nonumber\\
 & & \qquad -2\omega(r) r^2 \sin^2\theta d\phi dt\;,
 \end{eqnarray}
 where $\omega(r)$ is the angular velocity of a locally non-rotating frame
at radius $r$ measured by an asymptotic inertial observer. Under these
assumptions, the quantities $\nu(r)$ and $m(r)$ satisfy the
Oppenheimer-Volkoff equations
 \begin{equation}
 \frac{dP}{dr}=-(P+\rho)\frac{m+4\pi r^3P} 
   {r^2(1-2m/r)}
 \label{eqP}
 \end{equation}
and
 \begin{equation}
 \frac{dm}{dr}=4\pi r^2 \rho\;,
 \label{eqM}
 \end{equation}
 as well as the equation
 \begin{equation} 
 \frac{d\nu}{dr}=2\frac{m+4\pi r^3 P}
   {r^2(1-2m/r)}\;,
 \label{eqNu}
 \end{equation}
 where $P$ is the pressure and $\rho$ the energy density at a radius $r$
inside the neutron star.  In order to solve for the quantity $\omega(r)$,
and hence for the moment of inertia of the neutron star, we define,
following Hartle~\cite{H67},
 \begin{equation}
 f(r)\equiv 1- \frac{\omega(r)}{\Omega}
 \end{equation}
and
 \begin{equation}
 j\equiv e^{-\nu/2}\left(1-2\frac{m}{r}\right)^{1/2}\;,
 \end{equation}
 and solve the equation
 \begin{equation}
 \frac{d}{dr}\left(r^4 j \frac{df}{dr}\right)
   +4 r^3 f \frac{dj}{dr}=0
 \label{pdej}
 \end{equation}
 for the radial dependence of $f$. 

Given a fiducial density $\rho_0$ and the corresponding pressure $P_0$, as
well as a core mass $M_{\rm c}$ and radius $R_{\rm c}$, we intergrate
numerically Eqs.~(\ref{eqP}) and (\ref{eqM}) outwards from the
core-envelope interface until the intergration reaches the surface of the
neutron star defined by $P(R_{\rm NS})=0$. The mass of the neutron star is
then simply $m(R_{\rm NS})=M_{\rm NS}$. 

For the core configurations that maximize the neutron-star moment of
inertia (i.e., the one with constant energy density in the core), we
integrate analytically the same equations (Eqs.~[\ref{eqP}],
[\ref{eqM}]) but inwards from the
core-envelope interface. If the resulting neutron-star mass and radius
satisfy the constraints imposed by the kHz QPOs, we then solve numerically
Eq.~(\ref{eqNu}) with the boundary condition $\nu(R_{\rm
NS})=\ln\left[1-2M_{\rm NS}/R_{\rm NS} \right]$.

Equation~(\ref{pdej}) is a second-order partial differential equation with
the boundary conditions
 \begin{equation}
 \left[\frac{df}{dr}\right]_{r=0}=0
 \label{fin}
 \end{equation}
 and
 \begin{equation}
 f(r=R_{\rm NS}) = 1-2\frac{I}{R_{\rm NS}^3}\;,
 \label{fout}
 \end{equation}
 where $I$ is the moment of inertia of the neutron star given in closed
form by~\cite{RP94}
 \begin{eqnarray}
 I&=&-\frac{2}{3}\int_0^{R_{\rm
 NS}}\frac{dj(r)}{dr} f(r) r^3dr\nonumber\\
 &=&\frac{8\pi}{3}\int_0^{R_{\rm NS}}
 (\rho+P)\frac{f(r)j(r)r^4}{1-2m(r)/r}dr\;.
 \label{Ical}
 \end{eqnarray}
 We integrate Eq.~(\ref{pdej}) outwards from the center of the neutron
star, using boundary condition~(\ref{fin}) and a trial value $f_{\rm 0}$
at the inner boundary. We call this trial solution $f_{\rm tr}(r)$ and
calculate the corresponding moment of inertia $I_{\rm tr}$ using
Eq.~(\ref{Ical}). We note that Eq.~(\ref{pdej}) is scale-free
and therefore assuming a different value of $f$ at the inner boundary,
$f(r=0)=\xi f_0$, where $\xi$ is a constant, results in the solution
$f(r)=\xi f_{\rm tr}(r)$ and the corresponding moment of inertia $I=\xi
I_{\rm tr}$. Given the trial solution, we can then calculate the value of
the parameter $\xi$ for which the boundary condition~(\ref{fout}) is
satisfied, i.e.,
 \begin{equation}
 \xi=\left[f_{\rm tr}(r=R_{\rm NS})+
   \frac{2}{R_{\rm NS}^3} I_{\rm tr}\right]^{-1}
 \end{equation}
 As a result, the solution of Eq.~(\ref{pdej})
that satisfies boundary conditions~(\ref{fin}) and
(\ref{fout}) is just $f(r)=\xi f_{\rm tr}(r)$ and the
moment of inertia of the neutron star is
 \begin{equation}
 I=\frac{I_{\rm tr}}
 {f_{\rm tr}(r=R_{\rm NS})+ 2 I_{\rm tr}/ R_{\rm
   NS}^3}\;.
 \end{equation}

We performed all numerical integrations using a fourth-order Runge-Kutta
scheme. We verified our numerical implementation of the procedure outlined
above and our integration algorithms by taking the limit
$\rho_0~\rightarrow~\infty$, i.e., assuming that we know the equation of
state everywhere in the neutron star, and comparing the calculated moments
of inertia for different neutron star parameters and equations of state
with the values given by Cook, Shapiro, and Teukolsky~\cite{CST94}. We
performed an additional test by taking the limit $\rho_0\rightarrow 0$,
i.e., assuming a constant-density neutron star, and comparing the
calculated moments of inertia for different neutron star parameters with
the values given by Sabbadini and Hartle~\cite{SH77} and Abramowicz and
Wagoner~\cite{AW79}. In all tests the agreement was better than $\sim
0.5$\%.

\section{BOUNDS ON NEUTRON-STAR PROPERTIES}

\subsection{Optimal bounds on neutron-star masses and radii}

The constraints~(\ref{Mmax}) and (\ref{Rlim}) imposed on the neutron-star
mass and radius by the identification of the highest-frequency kHz QPO
with a Keplerian orbital frequency are derived without any assumptions
regarding the equation of state for neutron-star matter~\cite{MLP98}.
These constraints can be further optimized assuming an equation of state
for the envelope of the neutron star and constraining the properties of
the core (eqs.~[\ref{Mcmax}] and [\ref{Mcmin}]) so that they satisfy the
minimal set of assumptions discussed in \S2. For a given equation of state
and fiducial density $\rho_0$ and for each core mass and radius in the
allowed range defined by the limits~(\ref{Mcmax}) and (\ref{Mcmin}), we
integrate the Oppenheimer-Volkoff equations and calculate the resulting
neutron-star mass and radius.

Figure~1 shows the allowed ranges of stellar masses and radii for two
representative equations of state (labeled according to Ref.~\cite{CST94})
and for different values of the fiducial density $\rho_0$. In order to
understand the qualitative behavior of the bounds shown in Figure~1, for a
given equation of state, we call $\rho_c(M_{\rm NS})$ the central density
and $R_{\em eos}(M_{\rm NS})$ the radius of the star of mass $M_{\rm NS}$
for this specific equation of state.

\begin{figure}
\centerline
{\psfig{file=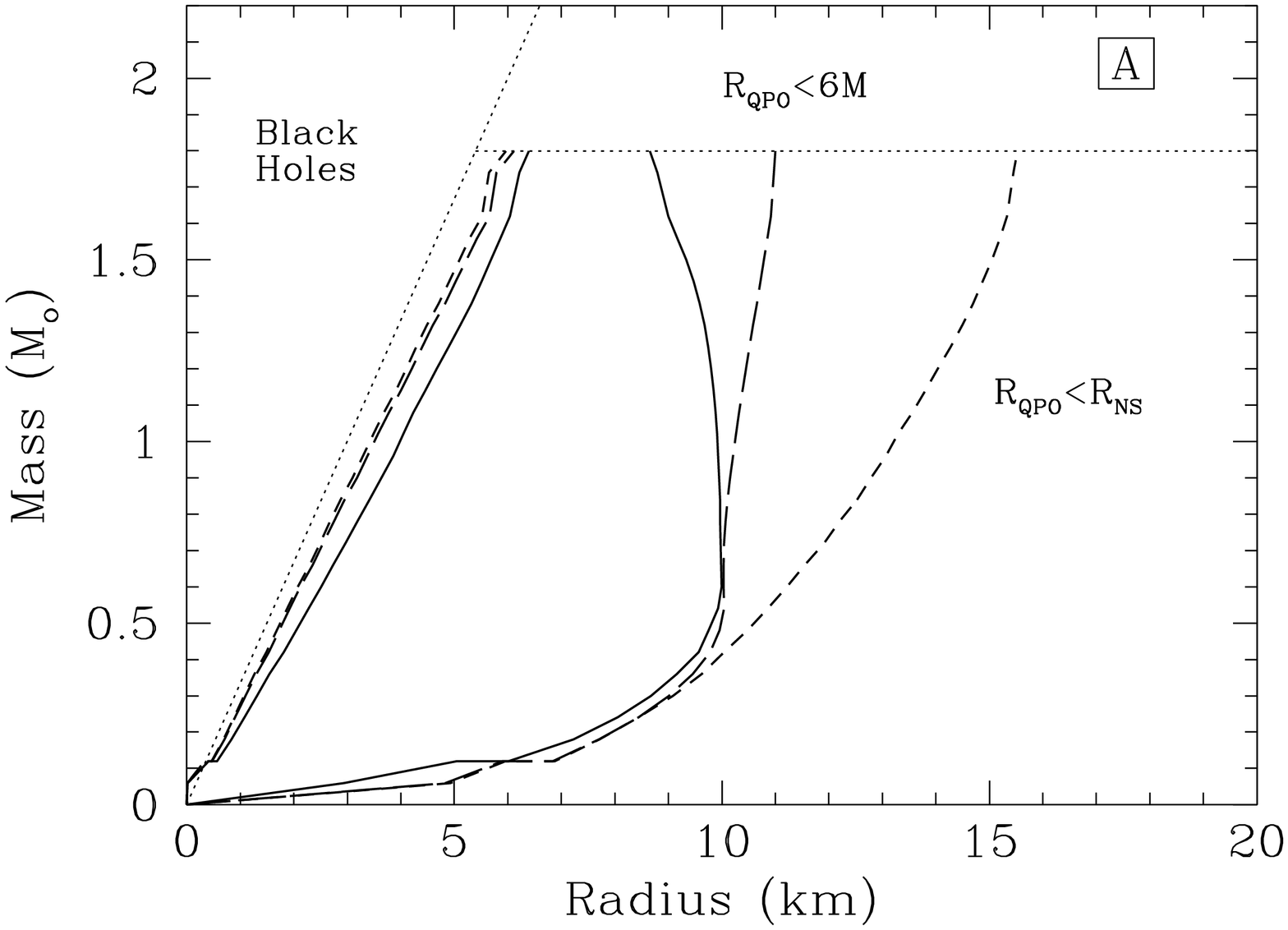,angle=-90,height=7truecm,width=9truecm}}
\centerline
{\psfig{file=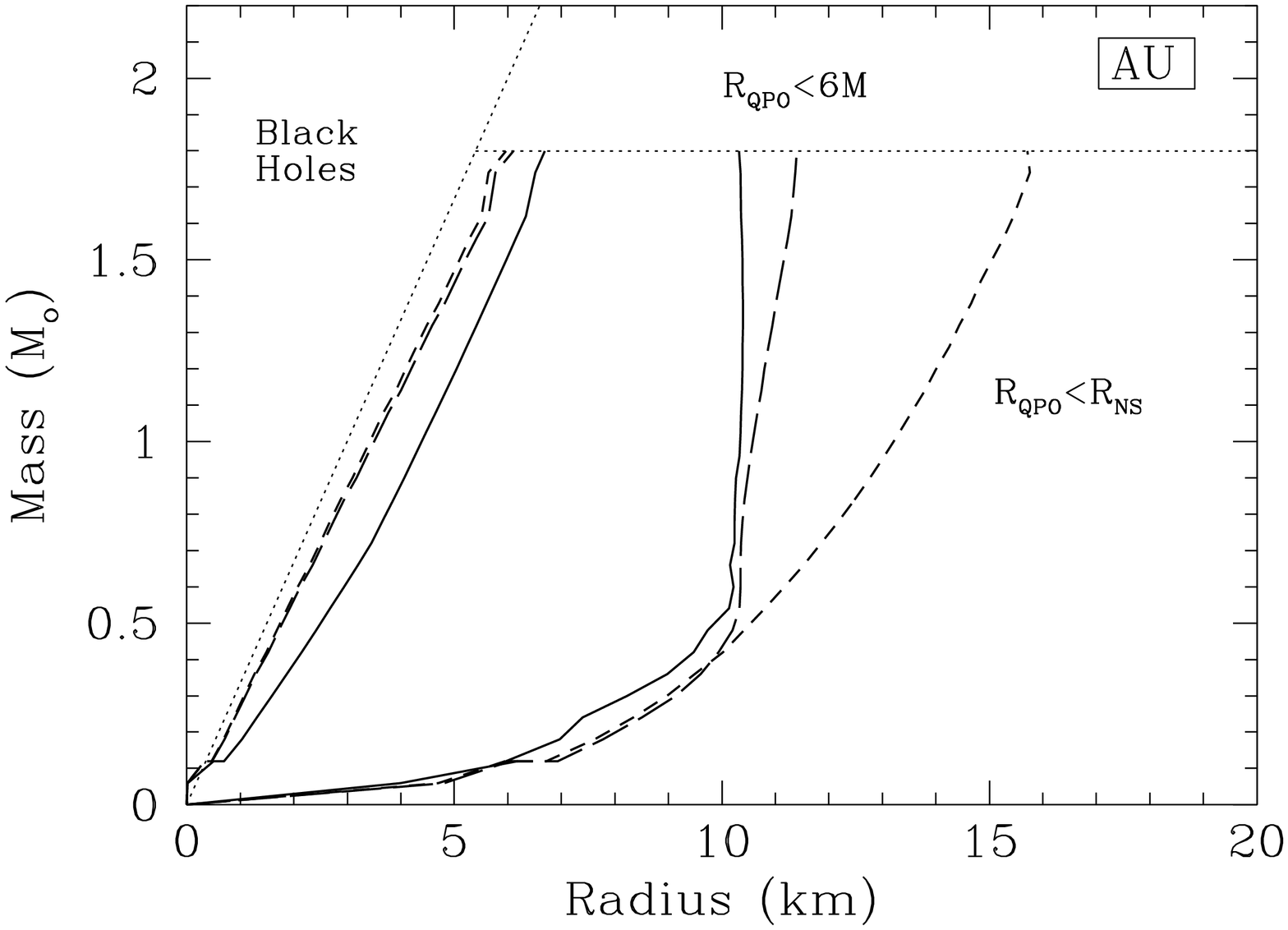,angle=-90,height=7truecm,width=9truecm}}
\caption
 {Bounds on the mass and radius of a neutron star imposed by the
identification of a 1220~Hz QPO with a Keplerian orbital frequency.
Different panels correspond to different equations of state, which were
assumed to be valid up to some fiducial density. Different line types
correspond to the constraints imposed for different values of the fiducial
density. {\em Dotted lines:\/} equation-of-state independent limits
(eqs.~[\ref{Mmax}] and [\ref{Rlim}]); {\em short-dashed lines:\/}
$\rho_0=2.7\times 10^{14}$~g\,cm$^{-3}$; {\em long-dashed lines:\/}
$\rho_0=7\times 10^{14}$~g\,cm$^{-3}$; {\em solid lines:\/}
$\rho_0=2\times 10^{15}$~g\,cm$^{-3}$. The dotted lines for the maximum
allowed radii in both panels overlap with the short-dashed lines and have
been omitted for clarity.}
 \end{figure}

In the limiting case $\rho_0\rightarrow 0$, i.e., when the equation of
state is assumed to be unknown everywhere in the star, the lower bound on
the neutron-star radius corresponds to the general relativistic
requirement $R_{\rm NS}\ge (9/4)M_{\rm NS}$~\cite{SH77}, so that the
central pressure in the neutron-star is not infinite. The upper bound on
the neutron-star radius corresponds to the bound~(\ref{Rlim}) set by the
kHz QPOs.

\begin{figure}
\centerline
{\psfig{file=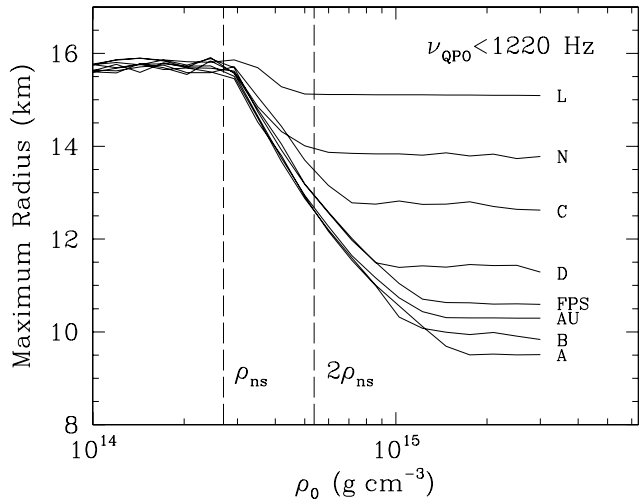,angle=-90,height=7truecm,width=9truecm}}
\caption
 {Maximum neutron-star radius allowed by the identification of a 1220~Hz
QPO with a Keplerian orbital frequency, as a function of the fiducial
density $\rho_0$ up to which an equation of state was assumed to be known.
Different curves correspond to different equations of state. The nuclear
saturation density is $\rho_{\rm ns}=2.7\times 10^{14}$~g\,cm$^{-3}$.}
 \end{figure}

For a non-zero fiducial density, the lower bound on the stellar radius
corresponds to the requirement $R_{\rm c}\ge (9/4) M_{\rm c}$ imposed on
the {\em core} (and not the stellar) properties. At high fiducial
densities, the mass and radius of the envelope become non-negligible,
$R_{\rm NS}>R_{\rm c}$, and hence the lower bound on the neutron-star
radius for a given mass becomes tighter (moves to the right in Figure 1).
For a non-zero fiducial density, the upper bound on the stellar radius
becomes tighter, as well. The radius of a star of given mass decreases
with increasing central density and hence, when $\rho_0>\rho_c(M_{\rm
NS})$, the maximum allowed radius of a neutron star with mass $M_{\rm NS}$
is $R_{\em eos}(M_{\rm NS})$, which may be smaller than the upper bound on
the radius imposed by the kHz QPOs.

In Figure~2 we plot the maximum allowed neutron-star radius as a function
of the fiducial density, for different equations of state (labels
according to Ref.~\cite{CST94}). As indicated from Figure 1 already, using
any equation of state up to about the nuclear saturation density
($\rho_{\rm ns}\simeq 2.7\times 10^{14}$~g\,cm$^{-3}$) does not not alter
the bounds on the neutron-star radius (eq.~[\ref{Rlim}]) imposed by the
kHz QPOs. Extending the use of any equation of state to higher densities
(up to two or four times $\rho_{\rm ns}$) leads to tighter constraints on
the stellar radius that depend strongly on the assumed equation of state.
The bound imposed by Eq.~(\ref{Mmax}) on the neutron star mass is not
affected by the knowledge of the equation of state.

Note here that all physical equations of state satisfy the set of minimal
assumptions used in constructing Figures~1 and 2. As a result, the bounds
on the mass and radius of a neutron star plotted in these figures will not
lead to any constraints on the microscopic properties of the equation of
state at high densities in addition to those already imposed by
relations~(\ref{Mmax}) and (\ref{Rlim}). However, these considerations do
impose additional constraints on the macroscopic properties of individual
neutron-stars, which in principle can be compared with other independent
estimates of their masses and radii. Such additional estimates may be
obtained from the X-ray spectra~\cite{LPT96} or the amplitudes of the
coherent oscillations during X-ray bursts~\cite{S92}.

\subsection{Bounds on neutron-star moments of inertia}

For a given equation of state and fiducial density $\rho_0$ and for a
neutron star with mass and radius consistent with the bounds calculated
above, we calculate the upper bound on the stellar moment of inertia, as
outlined in \S2. Figure~3 shows the maximum value of the quantity
$I_{45}/(M_{\rm NS}/M_\odot)$, where $I\equiv I_{45} 10^{45}$~g\,cm$^{3}$
is the neutron-star moment of inertia, as a function of the fiducial
density for different equations of state and for a maximum kHz QPO
frequency of 1220~Hz. The maximum ratio $I_{45}/(M_{\rm NS}/M_\odot)$
decreases with increasing fiducial density $\rho_0$, in part because
$I/M_{\rm NS}\propto R_{\rm NS}^2$ and the maximum allowed neutron-star
radius decreases with increasing $\rho_0$ (see Figure~2). For fiducial
densities higher than the nuclear saturation density, the maximum ratio
$I_{45}/(M_{\rm NS}/M_\odot)$ depends on the assumed equation of state.

For a neutron-star showing kHz QPOs with maximum frequency $\nu_{\max}$,
we can obtain the maximum value of the ratio $I_{45}/(M_{\rm
NS}/M_\odot)$, independent of the equation of state, by taking the limit
$\rho_0\rightarrow 0$. This corresponds to a star with constant density
and is the configuration that

\begin{figure}
\centerline
{\psfig{file=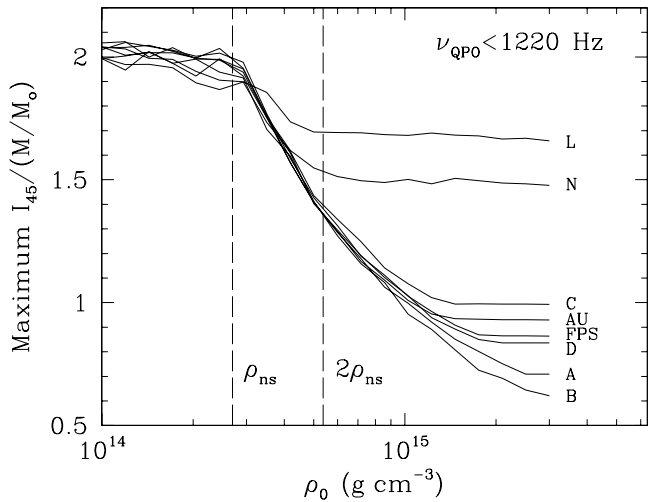,angle=-90,height=7truecm,width=9truecm}}
\caption
 {Maximum neutron-star moment of inertia in units of its mass allowed by
the identification of a 1220~Hz QPO with a Keplerian orbital frequency, as
a function of the fiducial density $\rho_0$ up to which the equation of
state was assumed to be known. Different curves correspond to different
equations of state. The nuclear saturation density is $\rho_{\rm
ns}=2.7\times 10^{14}$~g\,cm$^{-3}$.}
 \end{figure}   

\begin{figure}
\centerline
{\psfig{file=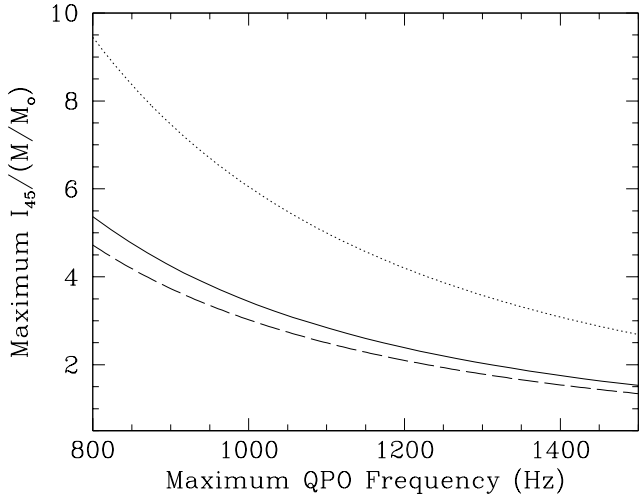,angle=-90,height=7truecm,width=9truecm}}
\caption
 {Maximum neutron-star moment of inertia in units of its mass as a
function of the maximum observed kHz QPO frequency. The upper bounds
calculated using the numerical method described in \S2 ({\em solid line})
are compared to the ones calculated analytically in the limit of zero
({\em dashed line}) and maximum ({\em dotted line}) compactness of the
neutron star.}
 \end{figure}

\noindent maximizes its moment of inertia. Figure~4 (solid line)
shows the resulting dependence; for a maximum kHz QPO frequency of
1220~Hz, $I_{45}/(M_{\rm NS}/M_\odot)\lesssim 2.3$.

We can obtain the dependence of the maximum ratio $I_{45}/(M_{\rm
NS}/M_\odot)$ on $\nu_{\rm max}$ at the limit $\rho_0\rightarrow 0$ in the
following {\em analytical} way. For any star, independent of the equation
of state, Sabbadini and Hartle~\cite{SH77} showed that
 \begin{equation}
 I\le \xi M_{\rm NS} R_{\rm NS}^2\;.
 \label{Imax}
 \end{equation}
 The parameter $\xi$ is a function of the compactness of the star $(M_{\rm
NS}/R_{\rm NS})$ and has the limiting values $\xi_{\rm min}=2/5$ in the
Newtonian limit ($M_{\rm NS}/R_{\rm NS}\rightarrow 0$) and $\xi_{\rm
max}=0.799\simeq 4/5$ in the limit of maximum compactness allowed by
general relativity (i.e., $M_{\rm NS}/R_{\rm NS}=4/9$)~\cite{SH77}.
Combining equation~(\ref{Imax}) with the radius bounds~(\ref{Mmax}) and
(\ref{Rlim}), we obtain
 \begin{equation}
 \frac{I}{M_{\rm NS}} \le \frac{\xi}
   {24\pi^2\nu_{\rm max}^2}
 \le\frac{1}
   {30\pi^2\nu_{\rm max}^2}\;,
 \label{IoMa}
  \end{equation}
 where $\xi$ corresponds to the compactness of the neutron star that
maximizes the ratio $I/M_{\rm NS}$ and in the last inequality we used the
fact that $\xi\le 4/5$. In Figure~4 we compare the analytical upper
bound~(\ref{IoMa}) for the two limiting values of $\xi$ ($\xi=2/5$: dashed
line; $\xi=4/5$: dotted line) to the one calculated numerically in the
limit $\rho_0\rightarrow 0$ (solid line).

When $\rho_0\rightarrow 0$, the maximum value of the ratio $I/M_{\rm NS}$
corresponds to a neutron star of constant density with typically the
maximum radius allowed by constraint~(\ref{Rlim}) imposed by the kHz QPOs.
For such a neutron star, $M_{\rm NS}/R_{\rm NS}=1/6$ (see
Eqs.~[\ref{Mmax}] and [\ref{Rlim}]) and hence $\xi$ is closer to the
Newtonian limit of $2/5$ than to its maximum value of $4/5$, as suggested
by Figure~4. Indeed, by comparing the analytical scaling to the numerical
result we find that a better
approximation to the upper bound of the ratio $I_{45}/(M_{\rm
NS}/M_\odot)$ corresponds to $\xi=0.452$ and is (see also Figure~4)
 \begin{equation}
 \frac{I_{45}}{M_{\rm NS}/M_\odot} \lesssim
   2.3 \left(\frac{1220~\mbox{Hz}}{\nu_{\rm max}}
     \right)^2\;.
 \end{equation}
 This is the maximum value of the ratio $I_{45}/(M_{\rm NS}/M_\odot)$
allowed by general relativity for a neutron star that shows a maximum kHz
QPO frequency of $\nu_{\rm max}$ in its power spectrum.

\subsection{Maximum nodal precession frequency}

The orbital plane of an infinitesimally inclined circular orbit around a
slowly spinning neutron star precesses because of general relativistic
frame dragging, as well as because of classical effects related to the
quadrupole moment of the stellar gravitation field. The nodal precession
frequency $\nu_{\rm NP}$ of such an orbit is~\cite{SV98}
 \begin{equation}
  \nu_{\rm NP}=\nu_{\rm LT}-\nu_{\rm C}=
  \frac{8\pi^2 I}{M_{\rm NS}}\,\nu_{s}\,
  \nu_{\rm K}^2\,-\,\nu_{\rm C}\;,
 \label{nugen}
 \end{equation}
 where $\nu_{\rm LT}$ and $\nu_{\rm C}$ are the contributions of the
general relativistic Lense-Thirring precession and of the classical
precession, respectively, $\nu_{\rm s}$ is the spin frequency of the
neutron star, and $\nu_{\rm K}$ is the Keplerian orbital frequency of the
orbit. 

The frequency of the general relativistic precession is directly
proportional to the ratio $I/M$, for which we have obtained optimal bounds
in the previous section. Given that the effect of classical precession is
to reduce the nodal precession frequency and using the upper
bound~(\ref{IoMa}) on the ratio $I/M_{\rm NS}$ we obtain
 \begin{equation}
 \nu_{\rm NP} \le \frac{1}{3}\xi\nu_{\rm s}
 \simeq 45.2\left(\frac{\nu_{\rm s}}
   {300~\mbox{Hz}}\right)~\mbox{Hz}\;,
 \label{numax}
 \end{equation}
 where we used $\xi\simeq 0.452$, for the reasons
discussed in \S3.2.

Equation~(\ref{numax}) shows that there exists an upper limit on the nodal
precession frequency of a circular orbit around a slowly spinning neutron
star, which we have obtained analytically. It is remarkable that this
upper limit is independent of all the other properties of the neutron
star, of the unknown equation of state, or of the properties of the
circular orbit.

\section{DISCUSSION}

We have calculated optimal bounds on the masses, radii, and moments of
inertia for slowly rotating neutron star in which observed frequencies of
kHz QPOs have been identified with Keplerian orbital frequencies.  
Assuming the validity of an equation of state up to some fiducial density
$\rho_0$, these bounds become tighter as $\rho_0$ increases. In the
limiting case of $\rho_0\rightarrow 0$, i.e., when we make no assumption
regarding the equation of state, we have derived an analytical upper bound
on the neutron-star moment of inertia for a given maximum observed kHz QPO
frequency. We also obtain analytically the maximum nodal precession
frequency of an inclined circular orbit around a neutron star, which
depends {\em only\/} on the spin frequency of the star and is independent
of the other stellar properties, the equation of state, or the properties
of the circular orbit.

In this section we use these constraints to address the possible
observational evidence for general relativistic frame-dragging effects in
the rapid variability of accreting neutron stars~\cite{SV98}. Many such
sources often show three distinct types of QPOs that are not harmonically
related~\cite{vdk98,MLP98}. The two QPOs at kilohertz frequencies are
believed to occur at the Keplerian frequency of a stable circular orbit in
the accretion disk and at its beat with the neutron-star spin
frequency~\cite{Setal96,MLP98}. According to Stella \& Vietri~\cite{SV98},
the third, low-frequency QPO (the HBO) occurs at the nodal precession
frequency of the orbit responsible for the kHz QPO.

We can first test quantitatively the suggestion that the HBO occurs at the
nodal precession frequency of an inclined circular orbit by comparing the
maximum observed HBO frequency in different sources with the maximum
possible nodal precession frequency around a slowly spinning neutron star
(Eq.~[\ref{numax}]). We will use the data of five bright neutron stars, in
which all three QPOs can be identified unambiguously, as discussed in
detail in Ref.~\cite{Petal98}. Figure~5 shows the maximum HBO frequencies
in these sources plotted against their spin frequencies inferred from the
peak separation of the kHz QPOs. The data points are compared to the
maximum possible nodal precession frequency around a neutron star
calculated in \S\,3.2 (Eq.~[\ref{numax}]).  Four out of the five sources
are inconsistent with our optimal bound at least at the one-sigma level.
In particular, the observed HBO frequencies in the source GX~17$+$2 can be
excluded at high statistical significance from being nodal precession
frequencies, if the peak separation of the kHz QPOs is equal to the spin
frequency of the neutron star.

It is possible that the observed HBO occurs at the second harmonic of the
nodal precession frequency. A precessing circular orbit has a two-fold
symmetry that could, in principle, produce even-order harmonics that are
stronger than the odd-order harmonics. The observation of a subharmonic of
the HBO in neutron-star sources as well as of subharmonics of similar QPOs
in black-hole sources~\cite{FK98} gives additional weight to this
conjecture. In this case, we address the nodal-precession interpretation
of the HBO using the observed correlation between the HBO and kHz QPO
frequencies in the same five bright neutron-star sources discussed in
Ref.~\cite{Petal98}.

When the effects of classical precession are negligible, observation of a
Keplerian, spin, and nodal precession frequencies leads to a direct
measurement of the ratio  

\begin{figure}
\centerline
{\psfig{file=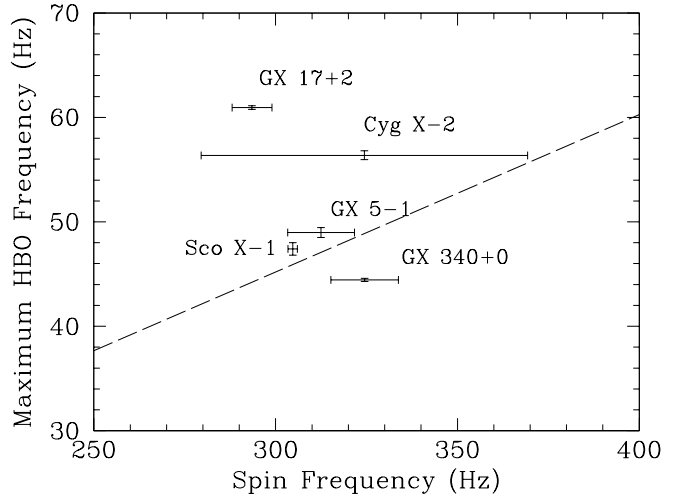,angle=-90,height=7truecm,width=9truecm}}
\caption
 {Maximum HBO frequency versus inferred spin frequency for five bright
neutron-star sources. The dashed line is the maximum nodal precession
frequency for an inclined circular orbit around a star of a given spin
frequency.}
 \end{figure}

\noindent $I/M$ of the neutron star (see
Eq.~[\ref{nugen}]). Identification of the HBO frequency in four of these
five sources with the second harmonic of the nodal precession frequency
requires $I_{45}/(M_{\rm NS}/M_\odot)\ge 2.3$ at the 99\% confidence
level~\cite{Petal98}. However, for the
maximum observed kHz QPO frequency of 1220~Hz, $I_{45}/(M_{\rm
NS}/M_\odot)\le 2.3$, independent of the equation of state (extreme case
of $\rho_0=0$).  Moreover, assuming even the unrealistically stiff
equation of state L up to twice nuclear saturation density results in an
upper limit of $I_{45}/(M_{\rm NS}/M_\odot)\le 1.7$ (see Figure~3).
Assuming any other equation of state results in even lower values of the
ratio $I_{45}/(M_{\rm NS}/M_\odot)$ which are inconsistent at high
statistical significance with the value required by the nodal-precession
interpretation of the HBO.

We therefore conclude that the nodal-precession interpretation of the HBO
observed in several neutron-star sources is inconsistent with the
identification of the higher-frequency kHz QPO with a Keplerian frequency
of a circular orbit and with the identification of the frequency
separation of kHz QPOs with the spin frequency of the neutron
star~\cite{SV99}.

\acknowledgments
 We are grateful to Greg Cook and Cole Miller for providing us with the
pressure-density relations of the various equations of state for
neutron-star matter. We thank the referee R.\ Wagoner for his prompt and
helpful responce. We also thank for their hospitality the astronomy group
at the University of Leicester, the astronomical institute of the
University of Amsterdam, and the Max-Planck Institute for Radioastronomie
in Bonn, where parts of this work were completed. This work was supported
in part by post-doctoral fellowships of the Smithsonian Institute.

\end{document}